\documentclass[a4paper,10pt]{article}
\usepackage[geometry]{ifsym}
\usepackage[english]{babel}
\usepackage{color}
\usepackage{soul}
\usepackage{mathrsfs}
\usepackage{amsfonts,amsmath,latexsym}
\usepackage{tabularx,ragged2e,booktabs,caption}
\newcolumntype{C}[1]{>{\Centering}m{#1}}

\usepackage{enumerate}
\usepackage{dsfont}
\usepackage[T1]{fontenc}
\usepackage{graphicx}
\usepackage{enumerate}
\usepackage{epsf}
\usepackage{amsthm}
\usepackage{url}
\usepackage{amssymb}
\usepackage{fancyhdr} 
\usepackage{palatino}

\def\R{\mathbb R}
\def\N{\mathbb N}

\def\E{\mathbb E}
\def\I{\mathbb I}

\def\S{\mathbb S}

\usepackage{mathptmx}
 
\author{
{\sc}\ {\sc Jia Liu}
\thanks{Corresponding author, Department of Mathematics and Statistics,  University of Jyv\"askyl\"a, P.O.Box (MaD) FI-40014  Finland e-mail:{\tt  jia.liu@jyu.fi}} 
}

\date{}
\title{An improved EM algorithm for solving MLE in constrained diffusion kurtosis imaging of human brain}
\newcommand{\MBFigure}[6]{
$\left. \right.$ \\
\refstepcounter{figure}
\addcontentsline{lof}{figure}{\numberline{\thefigure}{\ignorespaces #5}}
\begin{center}
\begin{minipage}{#1cm}
\centerline{\includegraphics[width=#2cm,angle=#3]{#4}}
\begin{center}
\upshape{F\textsc{ig} \normal
\end{center}
size{\thefigure}. $-$} #5
\end{center}
\label{#6}
\end{minipage}
\end{center}
$\left. \right.$ \\}
 
\begin{document}
\maketitle 

\begin{abstract}
The displacement distribution of a water molecular is characterized mathematically as Gaussianity without considering potential diffusion barriers and compartments. However, this is not true in real scenario: most biological tissues are comprised of cell membranes, various intracellular and extracellular spaces, and of other compartments, where the water diffusion is referred to have a non-Gaussian distribution. Diffusion kurtosis imaging (DKI), recently considered to be one sensitive biomarker, is an extension of diffusion tensor imaging, which quantifies the degree of non-Gaussianity of the diffusion. This work proposes an efficient scheme of maximum likelihood estimation (MLE) in DKI: we start from the Rician noise model of the signal intensities. By augmenting a Von-Mises distributed latent phase variable, the Rician likelihood is transformed to a tractable joint density without loss of generality. A fast computational method, an expectation-maximization (EM) algorithm for MLE is proposed in DKI. To guarantee the physical relevance of the diffusion kurtosis we apply the ternary quartic (TQ) parametrization to utilize its positivity, which imposes the upper bound to the kurtosis.  A Fisher-scoring method is used for achieving fast convergence of the individual diffusion compartments. In addition, we use the barrier method to constrain the lower bound to the kurtosis. The proposed estimation scheme is conducted on both synthetic and real data with an objective of healthy human brain. We compared the method with the other popular ones with promising performance shown in the results.
\end{abstract}


\section{Introduction}
\label{sec:intro}
Magnetic resonance (MR) is capable of measuring the displacement diffusion of water molecules and provides a unique insight into image contrasts reflecting anatomical architectures inside organic tissues. Diffusion tensor imaging (DTI) is one of the noninvasive imaging modalities based on the diffusion weighted (DW-) MR measurements. It captures the neurostructural information by means of diffusion tensors, where the probability of the water diffusion is simply assumed to be Gaussian. However, this assumption is argued to diverge significantly from the genuine in many biological tissues, especially in human brain containing an appendage of complex microstructural-rich tissues, i.e. cell membranes, boundaries and other complex compartments, where the displacement distribution is no longer Gaussian.

Diffusion kurtosis imaging (DKI) is recently referred as a natural extension of DTI \cite{ghosh2014,jensen2005,veraart2011} and as one of high angular resolution diffusion imaging (HARDI, \cite{alexander2005,jayachandra2008}) techniques. It attempts to quantify the degree of diffusional deviation from the Gaussian density expressed by \cite{jensen2005,veraart2011,qi2009,steven2014,tabesh2011}
\begin{align}\label{DKI}
S(b) = S_0 \exp(-b D_{app} + \frac 1 {6} b^2 D_{app}^2 K_{app}),
\end{align}
where $b$ is the diffusion weighting amplitudes or so-called $b$ value, $D_{app}:= {\bf g}^T D {\bf g} = \sum \limits_{\ell_1,\ell_2 =1}^3 g_{\ell_1}g_{\ell_2} D_{\ell_1,\ell_2}$ is called the apparent diffusional coefficient, and $K_{app}$ is the apparent diffusion kurtosis with the further derivation
\begin{align*}
K_{app} = \biggl(\frac{ \overline{tr(D)} }{D_{app}}\biggr)^2  \sum \limits_{\ell_1,\ell_2,\ell_3,\ell_4 =1}^3 g_{\ell_1}g_{\ell_2} g_{\ell_3} g_{\ell_4} W_{\ell_1,\ell_2, \ell_3,\ell_4},
\end{align*}
where "$tr$" denotes the trace of the matrix operator, and $\overline{tr(D)} = \sum\limits_{i=1}^3 tr(D)$. The definition of kurtosis tensor $W_{\ell_1,\ell_2, \ell_3,\ell_4}$ can be found in \cite{jensen2005}. We follow \cite{ghosh2014} and \cite{veraart2011} and list three constraints in DKI:
\begin{enumerate}
\item[\# 1.] The physical relevance and biological plausibility require that $D$ is positive definite.
\item[\# 2.] $K_{app} > 0$ is the lower bound constraint on the apparent diffusion kurtosis, although in theory  $K_{app}\geq - 2$. This lower bound is in agreement with higher order ($\geq 4$) tensors in HARDI, depicting the complex structural information of fibers in the brain. It further implies that the fourth symmetric kurtosis tensor $W$ should be positive definite in three dimension (3d). 
\item[\# 3.]  The upper bound constraint is $K_{app} \leq 3/ (b D_{app})$. This limit is derived from the assumption that the signal intensity $S(b)$ is a monotonically decreasing function of the $b$-amplitudes. In other words, DKI can utilize $b$-values only less than 3000 $s/mm^2$, which however is much more feasible in clinic imaging protocols.   
\end{enumerate}
This paper has fourfold contributions: 
1) We use Von Mises data augmentation to transform the non-linear Rician likelihood into the joint likelihood in the general linear framework. This strategy provides a possibility to use the original Rician noise model in MRI with dramatically reduced the computational burden. 2) We propose a fast computational scheme for MLE in DKI by the EM algorithm. 3) The three constraints of the kurtosis tensor are explicitly adopted into the modeling, where we apply the ternary quartic (TQ) theory to guarantee the positivity of the kurtosis tensor with new parametrization. 4) We apply the barrier method combining with the Fisher scoring algorithm in DKI to complete the constraint \#3.
\section{Theory}
\subsection{MR noise and Rician magnitude}
We first recall the noise $\epsilon$ in the raw MR-acquisitions which is composed of two $i.i.d.$ Gaussian random variables, $\epsilon_r$ and $\epsilon_i$, with zero mean, and variance $\sigma^2$ specified from the real and imaginary components, respectively. The joint density of the MR noise is expressed by $
 p_ {S, \sigma^2}(\epsilon_r, ~ \epsilon_i )= \frac 1 {2\pi \sigma^2} 
 \exp\biggl( - \frac {\epsilon_r^2 + \epsilon_i^2 } {2\sigma^2} \biggr)$.   
The magnitude $Y$ of the MR signal, as a consequence, is Rician distributed with the likelihood function
\begin{align}\label{(1)}
p_ {S, \sigma^2}( y ) = \frac{ y }{\sigma^2} 
\exp\biggl(  - \frac{ y^2+S^2 }{2\sigma^2} \biggr)
I_0\biggl(  \frac{ y S }{\sigma^2}\biggr) \mathds{1} (S \geq 0) , 
\end{align}
where $S$ denotes the signal intensity corrupted by the complex valued noise having the magnitude $Y=|S+ \epsilon| = \sqrt{(S+\epsilon_r)^2 + \epsilon_i^2 }$,  $I_{\alpha}(\cdot)$ is the $\alpha$-order modified Bessel function of the first kind, and $  \mathds{1} (\cdot) $ is the indicator function.
\subsection{Von Mises augmentation}
Let $ \varphi$ be the phase data defined as $  \varphi:= \arg\biggl(   S + \epsilon_r + \mathbf{i}  \epsilon_i \biggr)  \in [0,2\pi)$ such that $ S + \epsilon_r =Y\cos(\varphi)$ and $  \epsilon_i  =Y \sin(\varphi)$. By the Jacobian transformation, the joint density of 
$\varphi$ and $Y$ with the parameters $ S $ and $\sigma^2$ is 
\begin{align} \label{joinlik}
 p_{S,\sigma^2}( y, \varphi )&= \frac y {2\pi \sigma^2} 
 \exp\biggl(  - \frac 1 {2\sigma^2} \bigl(   y  \cos( \varphi) - S )^2  - \frac 1 {2\sigma^2}  y^2 \sin( \varphi)^2 \biggr)   &\notag \\ &
 =\frac y {2\pi \sigma^2}
\exp\biggl(  - \frac 1 {2 \sigma^2} \bigl( y^2 + S^2 - 2 S y \cos(\varphi) \bigr) \biggr)  
 = p_{S,\sigma^2} ( y) p_{S,\sigma^2}( \varphi |y  ),
 \end{align}
 where the  conditional density
\begin{align}\label{vonmises} 
  p_{S,\sigma^2}( \varphi |y  )  =  \frac 1{ 2\pi    I_0( S y/\sigma^2) }  
\exp\biggl( \frac{ S  y }{\sigma^2} \cos( \varphi ) \biggr)  , \quad \varphi \in [0,2\pi),
\end{align}
is an instance of the Von Mises distribution on the unit circle  symmetric around zero.  Note that although in theory the zero magnitude is obtained with zero probability density, in practice, we can still acquire zero measurements after discretization by the scanner. In such a case the MR noise contains only the real Gaussian component and the data has a Gaussian likelihood  $p_{S,\sigma^2} ( \epsilon_r=-S,~\epsilon_i=0  )=  \frac 1 {2\pi \sigma^{2} } \exp\biggl( - \frac {S^2} { 2\sigma^2}\biggr).$

\subsection{DTI and DKI} 
Under the typical assumption of Gaussian approximation of the diffusion displacement density of water molecules, the DTI signal model can be expressed in the form 
$S(b) = S_0 \exp(-b D_{app}^{(n)})$
with parametrization
$-b D_{app}^{(n)} =  Z \theta$ by
\begin{align}\label{DTI} 
S  = S_0\exp(Z(b, {\bf g}) \theta),
\end{align}
where $D_{app}^{(n)} :=\sum\limits_{\ell_1 =1}^3 \sum\limits_{\ell_2 =1}^3 \cdots \sum\limits_{\ell_{n} =1}^3  D_{\ell_1, \ell_2, \dots ,\ell_{n}} 
 g_{\ell_1} g_{\ell_2}\cdots g_{\ell_{n}}  \;$ with even number $n\in \N$. The tensor parameter is denoted by $\theta$ and $Z$ is a design matrix.
For a rank-2 DTI model, the six distinct elements of $D$ are defined as the vector parameter\\ $\theta_D = (\theta_1,\dots, \theta_6)^{\top}:=  
\bigl( D_{11}, D_{22} , D_{33}, D_{12}, D_{13}, D_{23} \bigr)^{\top}.$ The corresponding design matrix, composed of 
$m$ acquisitions is given by
\begin{align}  \label{matrx:Z_D}
Z_D= Z(b, {\bf g})  = -b
 \begin{pmatrix}
g_{11}^2& g_{21}^2 &   g_{31}^2 & 2 g_{11} g_{21}& 2 g_{11}g_{31}& 2 g_{21} g_{31}  \\
  \vdots  & \vdots  & \vdots & \vdots & \vdots & \vdots \\
g_{1j}^2& g_{2j}^2 &   g_{3j}^2 & 2 g_{1j} g_{2j}& 2 g_{1j}g_{3j}& 2 g_{2j} g_{3j}\\
\vdots  & \vdots  & \vdots & \vdots& \vdots & \vdots  \\
g_{1m}^2& g_{2m}^2 &   g_{3m}^2 & 2 g_{1m} g_{2m}& 2 g_{1m}g_{3m}& 2 g_{2m} g_{3m}
\end{pmatrix}. 
\end{align}   
\if
$Z_D= Z(b, {\bf g})  = -b
 \begin{pmatrix}
g_{11}^2& g_{12}^2 &   g_{13}^2 & 2 g_{11} g_{12}& 2 g_{11}g_{13}& 2 g_{12} g_{13}  \\
  \vdots  & \vdots  & \ddots & \vdots  \\
g_{i1}^2& g_{i2}^2 &   g_{i3}^2 & 2 g_{i1} g_{i2}& 2 g_{i1}g_{i3}& 2 g_{i2} g_{i3}\\
\vdots  & \vdots  & \ddots & \vdots  \\
g_{m1}^2& g_{m2}^2 &   g_{m3}^2 & 2 g_{m1} g_{m2}& 2 g_{m1}g_{m3}& 2 g_{m2} g_{m3}
 \end{pmatrix},$
\fi
With the new parametrization, the DKI model Eq.\eqref{DKI} can be further presented by
\begin{align}\label{eq:DKI1} 
S(b) =& S_0 \exp \biggl(  -b \sum \limits_{\ell_1,\ell_2 =1}^3 g_{\ell_1}g_{\ell_2} D_{\ell_1, \ell_2}
+ \frac{b^2}{6} (\sum \limits_{\ell_1 =1}^3 
\frac{  D_{\ell_1 \ell_1}  }{3} )^2
\sum \limits_{\ell_1,\ell_2,\ell_3,\ell_4 =1}^3 g_{\ell_1}g_{\ell_2} g_{\ell_3} g_{\ell_4} W_{\ell_1,\ell_2, \ell_3,\ell_4} \biggr)  \notag
&\\ &
= S_0 \exp( Z_D \theta_D + Z_W \theta_W(\overline{tr(D)}^2; W)),
\end{align} 
where the design matrix can be e.g. $\{Z_W\in \R^{m\times 15}: Z_ {W_j} = \frac{b^2}{6}  (g_{1j}^4, g_{2j}^4,   g_{3j}^4 , 6 g_{1j}^2 g_{2j}^2,\\ 6 g_{1j}^2g_{3j}^2, 6 g_{2j}^2 g_{3j}^2, 12 g_{1j}^2 g_{2j}g_{3j},12 g_{1j} g_{2j}^2g_{3j},12 g_{1j}g_{2j}g_{3j}^2,4g_{1j}^3 g_{2j}, 4g_{1j}^3 g_{3j},4g_{2j}^3g_{1j},\\4g_{2j}^3 g_{3j},4g_{3j}^3  g_{1j},4g_{3j}^3 g_{2j}), j=1\cdots m\} $.
\subsection{Constrained DKI and its reparametrization} 
Since $D$ is a $3\times 3$ symmetric positive definite matrix, it can be always written in terms of a product of two triangular matrices, $D=U U^T$  by Cholesky decomposition. Without changing the design matrix $Z_D$, we can write the tensor parameter $\theta_D$ as a function 
of L: $\theta_D(L) = (L_1^2,  L_2^2 + L_4^2, L_3^2+L_5^2+L_6^2, L_1L_4, L_1L_5, L_4L_5+L_2L_6),$
and $U$ is a $3\times 3$ lower triangular matrix 
$U=   \begin{pmatrix}
  L_1 &      &   \\
  L_4     &  L_2 &  \\
   L_5    &    L_6  & L_3          
 \end{pmatrix}$
constructed from the elements of $L$.
The Jacobian $\triangledown_L \theta_D$ is 
\begin{align}  \label{matrx:J_L}
J_L= \frac{\partial \theta_D} {\partial L_{j=1,\cdots,6}} 
= \begin{pmatrix}
  2L_1 &      &  &  &  &   \\
       & 2L_2 &  &2L_4  &  & \\
       &      &2L_3 & &2L_5&2L_6  \\
    L_4   & &   &L_1 &&\\
     L_5   & &   &  &L_1& \\
      & L_6&   &L_5&L_4&L_2
 \end{pmatrix}. 
\end{align} 
The constraints \#1 and \#2 in DKI (see page 2) require that $W_{app}$ should be non-negative. In DTI this positive constraint is typically solved by Hilbert's Theorem \cite{hilbert1888}, proving that any real valued positive function can be written as a sum of three squares of quadratic forms. For a rank-4 tensor, the widely used methods are based on the strategy of Ternary quartic (TQ). It turns out that the non-negative TQ's of a non-negative 3d kurtosis tensor have an expression
\begin{align}
W_{app} = \sum \limits_{i=1}^3 \biggl({\bf v^T} q_i \biggr)^2 = {\bf v^T}QQ^T {\bf v} = {\bf v^T}G {\bf v},
\end{align}
where ${\bf v} = [g_1^2, g_2^2, g_3^2, g_1g_2, g_1g_3, g_2g_3]^T$, and $Q = [q_1| q_2| q_3]$ is a $6 \times 3$ matrix, containing three $6\times 1$ vectors $q_i$. The Gram matrix $G = QQ^T$ is a $6 \times6$ positive symmetric matrix composed of all fifteen kurtosis tensor elements plus six free parameters (see \cite{barmpoutis2011} for details). 
Let $\theta_Q := \overline{tr( D)} \begin{pmatrix}   q_1\\ q_2\\ q_3   \\  \end{pmatrix},$ and  $P_j=\frac{b^2}{6} \begin{pmatrix}
  \bf v v^T &  &       \\
       &\bf v v^T &  \\       
     & & \bf v v^T   
 \end{pmatrix}$  
is an $18 \times 18 $ matrix at the signal acquisition $j$. Then Eq. \eqref{eq:DKI1} can be then written by
\begin{align}\label{eq:DKI2} 
S = &S_0 \sum \limits_{j=1}^m \exp \biggl ( Z_{D_j} \theta_D(L) +   \theta_Q^T P_j \theta_Q \biggr).
\end{align} 
 
\section{Maximum likelihood and weighted least squares methods with constraints}
\subsection{Constrained MLE by EM algorithm }
In the optimization of the likelihood, we employ the EM (Expectation - Maximization) algorithm for maximum likelihood estimation with constraints (CMLE) in DKI.
The theory of the EM algorithm can be found in textbooks, e.g. \cite{mclachlan2007}. It typically proceeds in two steps and shortens the computational complexity by using augmented data: in the E-step we calculate the expectation of the log likelihood w.r.t the conditional distribution of the latent variable given the observations, the other parameters having fixed values; in the M-step, we find the ML parameter of $S_0^2$ and $\sigma^2$ by maximizing the augmented joint log likelihood quantities.

For concreteness, in our data augmentation we are able to work with the joint logarithmic likelihood derived from Eq. \eqref{joinlik} and Eq. \eqref{eq:DKI2} under the Rician density of the signal data. After omitting the constant, the joint log-likelihood function is given by
\begin{align}& \label{eq:logjointlik_Q}
  m\log(\sigma^{-2}) -  \frac 1 {2\sigma^2} \sum_{j=1}^m  \biggl\{   Y_j^2 + S_0^2 \exp \biggl ( 2 Z_{D_j}\theta_D +   2 \theta_Q^T P_j \theta_Q \biggr) \notag &\\&  - 
2 \cos( \varphi_j) Y_j  S_0 \exp \biggl (Z_{D_j} \theta_D   +  \theta_Q^T P_j  \theta_Q \biggr) \biggr\},
\end{align}
by the EM algorithm for MLE in DKI. For simplifying the notations, we define ${\bf \zeta}_j^{(k)} =   \exp(  Z_{Dj} \theta_D ^{(k)}), {\bf \psi}_j^{(k)} = \exp \biggl(     (\theta_Q^{(k)})^T P_j   \theta_Q^{(k)} \biggr )$, and ${\bf\tau}_j^{(k)} =  Y_j \bigl\langle \cos( \varphi_j)\bigr \rangle^{(k)} $ with $<\cdot>$ being introduced as a shorthand for expectation.
\\
In the EM-iteration, given the current parameter estimates $(\theta_D^{(k)}, \theta_Q^{(k)}, S_0^{(k)},({\sigma^2})^{(k)})$, we update the conditional expectation of $\cos \phi$ w.r.t. the conditional Von Mises distribution of $\phi$ in Eq. \eqref{vonmises}  by \begin{align}\label{eq:cosphi_mle}
\bigl\langle \cos \varphi_j  \bigr \rangle^{(k)}   \leftarrow  \frac{ I_1\biggl(  Y_j S_0 ^{(k)} 
 {\bf\zeta}_j^{(k)} {\bf \psi}_j^{(k)} (\sigma^{-2})^{(k)}\biggr  )  }{ I_0\biggl( Y_j S_0 ^{(k)} {\bf \zeta}_j^{(k)} {\bf \psi}_j^{(k)} (\sigma^{-2})^{(k)}\biggr  ) }. 
\end{align}
This formula is fairly easy to obtain from the first moment of the Von Mises distribution.
\\
In the M-step we update $S_0^2$ and $\sigma^2$ by their modes with the recursions 
\begin{align}\label{eq:S0_mle}
   S_0^{(k+1)} &\leftarrow     \frac{ \sum_{j=1}^m  {\bf \tau}_j^{(k)} {\bf \zeta}_j^{(k)} {\bf \psi}_j^{(k)}}{\sum_{j=1}^m   ({\bf \zeta}_j^{(k)})^2 ({\bf\psi}_j^{(k)})^2 },
  \end{align} 
and
\begin{align}\label{eq:sigma_mle}
   (\sigma^2)^{(k+1)}  &\leftarrow  \frac {1}{2(m-1)} \sum_{j=1}^m  \biggl\{   Y_j^2 + (S_0^{(k)})^2  ({\bf \zeta}_j^{(k)})^2 ({\bf\psi}_j^{(k)})^2- 
2  S_0^{(k)}   {\bf \tau}_j^{(k)} {\bf \zeta}_j^{(k)} {\bf \psi}_j^{(k)} \biggr\},
  \end{align}  
where $m$ is the number of acquisitions in each voxel.

To find the optimal (the mode) of parameters $\theta_D$ and $\theta_Q$ we use the Laplace approximation for the joint likelihood w.r.t to $\theta_D$ and $\theta_Q$, \Big(the marginal pdf $\pi(y;\theta_D,\theta_Q)$\Big) , respectively, with Gaussian forms.
This can be conducted by applying the Fisher scoring method (also referred as Gauss Newton method) to minimize the objective function, the minus Eq. \eqref{eq:logjointlik_Q}, given by
\begin{align}&\label{eq:objective}
  f(\Theta): = \frac 1 {2\sigma^2} \sum_{j=1}^m  \biggl\{  S_0^2 \exp \biggl ( 2 Z_{D_j}\theta_D(L) +   2 \theta_Q^T P_j \theta_Q \biggr) \notag &\\&   - 
2 \cos( \varphi_j) Y_j  S_0 \exp \biggl (Z_{D_j} \theta_D(L)   +  \theta_Q^T P_j  \theta_Q \biggr) \biggr\}.
  \end{align} 
The essential difference between the Fisher scoring method and the traditional Newton's method is that we use the Fisher or the empirical Fisher information instead of the Hessian matrix. To imposed the constraints, we apply barrier method (see e.g. \cite{ruszczynski2006}).  
These can be achieved by using the MATLAB optimization toolbox, function \verb fmincon   with interior point algorithm, where the following statements are required in order to apply this function. 

For updating $\theta$, we first update $L$, formulating the optimization problem:
\begin{align}&\label{eq:sysD}
  \mbox{minimize}\qquad f(L) - \mu \sum_{j=1}^m \ln (\nu_j)  &\notag \\& 
  \mbox{subject to}\qquad g_j(L) - \nu_j = 0,  \qquad j=1, \cdots, m.  \qquad \nu_j \geq 0 &\notag \\& 
\mbox{with}\qquad g_j(L) =  (\theta_Q^T)  \biggl[\frac{6}{b^2} P_j \biggr] \theta_Q - \frac{3}{b} \biggl[-\frac{1}{b} Z_{D_j}\biggr] \theta_D(L)^{(k+1)}, \qquad  g_j(L) \leq 0,
\end{align} 
where the function $-\mu \ln (\nu_j)$ built a "barrier" close to the boundary of the cone $\R_+^m$ preventing $\nu_j$ being close to the boundary. The positive scalar $\mu$ is called barrier parameter which
should be decreasing at each iteration, and $\lambda$ is the Lagrangian multiplier.

The Fisher information, being the expectation of observed information matrix, is given by
\begin{align*}& 
\bigl\langle -\mathcal{J}(L^{(k)}) \bigr\rangle := \E \bigl[ H(L,\lambda)]= \E \bigl[\bigtriangledown^2 f(L) + \sum \limits_{j=1}^m \lambda_j \bigtriangledown^2 g_j(L) \bigr]&\\&
= \biggl[J_L^T (\bigtriangledown^2 f( \theta_D)) J_L +  \bigtriangledown f( \theta_D) \frac{\partial^2  \theta_D(L)}{\partial L_{k} \partial L_{h} } \biggr] + \sum \limits_{j=1}^m \lambda_j  M_{D_j} &\\&
= (\sigma^{-2})^{(k)}\sum\limits_{j=1}^m \biggl\{ J_L^T\biggl( 2 (S_0^{(k)})^2  ({\bf \zeta}_j^{(k)})^2 ({\bf\psi}_j^{(k)})^2  Z_{Dj}^TZ_{Dj} -  S_0^{(k)}   {\bf \tau}_j^{(k)} {\bf \zeta}_j^{(k)} {\bf \psi}_j^{(k)}   Z_{Dj}^TZ_{Dj}\biggr) J_L\biggr\}&\\&  +  \sum \limits_{j=1}^m \lambda_j  M_{D_j} ,
\end{align*}
with
\begin{align*}   
M_{D_j}  :=& \bigtriangledown^2 g_j(L) &\\&=
\frac 3 {b^2}\biggl[ Z_{Dj} \frac{\partial^2 \theta_{D}(L)}{\partial L_{k} \partial L_{h} } \biggr]
=  \frac 3 {b^2} \begin{pmatrix}
  2 Z_{1j} &      &  &  Z_{4j} &  Z_{5j} &   \\
       & 2 Z_{2j} &  &   &  & Z_{6j} \\
       &      &2 Z_{3j} & & &   \\
       &Z_{4j} &   &2 Z_{2j} &&Z_{6j}\\
     Z_{5j}   & Z_{6j}&   &  &2 Z_{3j}& \\
      &  &   & Z_{6j}& & 2 Z_{3j}
\end{pmatrix}.
\end{align*} 
The gradient ($\bigtriangledown f(L) \in \R^d$ ) of $f(L)$ at the current recursion is
\begin{align*}&
\bigtriangledown f(L^{(k))}  =  (\sigma^{-2})^{(k)} \sum \limits_{j=1}^m \biggl\{ (S_0^{(k)})^2  ({\bf \zeta}_j^{(k)})^2 ({\bf\psi}_j^{(k)})^2 J_L Z_{D_j}^T   - S_0^{(k)} {\bf \tau}_j^{(k)} {\bf \zeta}_j^{(k)} {\bf \psi}_j^{(k)} J_L Z_{D_j}^T  \biggr\}, &\\&
\mbox{and} \qquad
\bigtriangledown g(L^{(k)})= \frac 3 {b^2} Z_{Dj} J^{(k)}.
\end{align*} 
Note that this method works with single and multiple shells with different $b$ values, meaning that $b$ can be a scalar or a vector. In addition, before calculating the Fisher information, we use the regularization technique to smooth the Hessian matrix ({\it i.f.f.} it is singular)  by adding a scalar: $H(L^{(k)}, \lambda^{(k)}) \rightarrow H(L^{(k)}, \lambda^{(k)}) + \|\S(\theta^{(k)}), \lambda \| \I$, where $\I$ is an identical matrix with dimension $d \times d, \mbox{with}~ \theta_D \in \R^d$. We define the score $(\S(L^{(k)}  , \in \R^d, \lambda ) : = \bigtriangledown f(L^{(k))} + \sum\limits_{j=1}^m \bigtriangledown \lambda_j g(L^{(k)})$, and $\| \cdot \|$ is the norm operator. The norm $\|\S(\theta^{(k)}), \lambda \|$ is one optimal choice of the Levengerg-Marquart parameter \cite{yamashita2001rate} before calculating the Fisher information. It makes the algorithm much more stable by avoiding the singularity of the Fisher or empirical Fisher information. Moreover, the barrier parameter is implicitly inside the MATLAB solver when calling \verb fmincon , the interior point method, monitoring the decreasing situation is one stopping criteria of this method. 
Finally, we map back to $\theta_D$ by $\theta_D^{(k+1)} \leftarrow U^{(k+1)} (U^T)^{(k+1)} $. Detailed interpretation of the calculation can be found in Appendix \ref{ape:1}.

Using the same idea, we update $\theta_Q $ by solving the following Lagrangian of problem:
\begin{align}&\label{sys:objectiveQ}
  \mbox{minimize}\qquad f(\theta_Q) - \mu \sum_{j=1}^m \ln (\nu_j)  &\notag \\& 
  \mbox{subject to}\qquad g_j(\theta_Q) - \nu_j = 0,  \qquad j=1, \cdots, m.  \qquad \nu_j \geq 0 &\notag \\& 
\mbox{with}\qquad g_j(\theta_Q) = (\theta_Q^T)^{(k+1)} \biggl[\frac{6}{b^2} P_j \biggr] \theta_Q^{(k+1)} - \frac{3}{b} \biggl[-\frac{1}{b} Z_{D_j}\biggr] \theta_D( L), \qquad  g_j(\theta_Q) \leq 0.
  \end{align} 
For simplification, we use the same notations ($\lambda, \mu, \mbox{and}~ \nu$) of what are used for as the general parameters when talking about the barrier method in the work, which of course will change case by case. Also we should emphasize that the constrained functions $g(\cdot)$ are derived from $K_{app} \leq 3/(bD_{app})$. Particularly, in this case we use the empirical Fisher information (also referred as the observed information matrix), being equal to the minus of Hessian matrix, $H(\theta_Q, \lambda)$,
given by
\begin{align*}&
\mathcal{J}(\theta_Q^{(k)}) = -  \bigtriangledown^2 f(\theta_Q) - \sum \limits_{j=1}^m   \lambda_j  \bigtriangledown^2 g_j(\theta_Q) &\\&
= - (\sigma^{-2})^{(k)} \sum_{j=1}^m   \biggl\{8 (S_0^{(k)} )^2  ({\bf \zeta}_j^{(k)} )^2 ({\bf\psi}_j^{(k)})^2  \theta_Q^T P_j^T P_j \theta_Q + 2 (S_0^{(k)})^2  ({\bf \zeta}_j^{(k)})^2 ({\bf\psi}_j^{(k)})^2  P_j&\\&
- 4 S_0^{(k)}   {\bf \tau}_j^{(k)} {\bf \zeta}_j^{(k)}{\bf\psi}_j^{(k)} \theta_Q^T P_j^T P_j \theta_Q -2 S_0^{(k)}   {\bf \tau}_j^{(k)} {\bf \zeta}_j^{(k)}{\bf\psi}_j^{(k)} P_j \biggr \} - \sum \limits_{j=1}^m 2 \lambda_j   \biggl[\frac{6}{b^2} P_j \biggr]  &\\&
\mbox{and the gradient of $f(\theta_Q)$ is} &\\&
 \bigtriangledown f(\theta_Q) =   (\sigma^{-2})^{k} \sum_{j=1}^m   \biggl\{ 2(S_0^{(k)} )^2 ({\bf \zeta}_j^{(k)} )^2 ({\bf\psi}_j^{(k)})^2   P_j \theta_Q - 2S_0^{(k)} {\bf \tau}_j^{(k)} {\bf \zeta}_j^{(k)} {\bf \psi}_j^{(k)} P_j \theta_Q \biggr \}, &\\&
 \mbox{and} &\\&
 \bigtriangledown g(\theta_Q) =  2(\theta_Q^T)^{(k)} \biggl[\frac{6}{b^2} P_j \biggr]. 
\end{align*}
Again we use the regularization technique to smooth $H(\theta_Q, \lambda)$ before calculating $\mathcal{J}(\theta_Q^{(k)})$. Finally, we extract $\widehat \theta_W$ from the Gram matrix \cite{barmpoutis2007} by  $G= Q^TQ / \overline{tr(\widehat D)}^2$.
The treatment of the singularity of Fisher or empirical Fisher information $\bigl(\bigl\langle - \mathcal{J}(L) \bigr\rangle,~ \mathcal{J}(\theta_Q^{(k)}\bigr)$, respectively) and the detailed barrier method combing with the Fisher scoring are discussed in Appendix \ref{ape:2}.

To implement the proposed MLE scheme, we need to find the optimal $\widehat \Theta=(\theta_D(\widehat L)$, $\widehat \theta_Q)$ by the constrained Fisher scoring (CFS) methods at each iteration when updating $S_0$ and $\sigma^2$ by using Eq. \eqref{eq:S0_mle} and Eq. \eqref{eq:sigma_mle}, respectively. 
This can be done by vectorizing the score $\S(\Theta, \lambda) =\begin{pmatrix}
 \S(L,  \lambda)  \\ \S(\theta_Q, \lambda) \end{pmatrix}$ and presenting the Fisher information $\mathcal{J}(\Theta, \lambda)$ by a sparse matrix $\begin{pmatrix}
 \bigl\langle -\mathcal{J}(L, \lambda) \bigr\rangle  &   \\
     & \mathcal{J}(\theta_Q,\lambda)  \end{pmatrix}$. Considering objective data from the brain which usually contain millions of voxels, this means that the proposed scheme may yield heavy computation. In practise, it is possible to update $S_0$ and $\sigma^2$ by Eq. \eqref{eq:S0_mle} and Eq. \eqref{eq:sigma_mle} till the optimal had been found, then update $\Theta$. The algorithm will stop until the tolerance reached by monitoring the value of logarithmic likelihood calculated from Eq. \eqref{eq:logjointlik_Q}.

\subsection{Constrain weighted least square- \textbf{CWLS}} The weighted least squares (WLS) is a commonly used method in diffusion MRI, see e.g. \cite{Zhu2007,veraart2011} and \cite{ghosh2014}. 
Here we just impose the constraints and emphasize the problem solving under the proposed scheme. 

The objective function is constructed from Eq. \eqref{eq:DKI2} to minimize the gap between the observations and signal intensities, which is given by
\begin{align}& \label{eq:cwls}
f(\Theta(L, \theta_Q)) =\frac 1 2 \sum \limits_{j=1}^m  w_j  \biggl(\log Y_j - \log S_0 - Z_{D_j} \theta_D(L) -  \theta_Q P_j \theta_Q \biggr)^2  
\end{align}
with the same constrained functions mentioned before. The choice of weights are free, some possibilities including $Y_j^2$, $S_j^2$ or $S_j^2 / S_0^2$, etc.   . In this work, we choose weights, $w_j = Y_j^2 /S_0^2$ and use the fixed values of $S_0$ from the WLS. 
The Hessian matrices w.r.t to $L$ and $\theta_Q$ are, respectively,
\begin{align*}&
H_{cwls}(L,\lambda) =   J_L^T \biggl(\sum \limits_{j=1}^m  w_j  \biggl(\log Y_j - \log S_0 - Z_{D_j} \theta_D(L) -  \theta_Q P_j \theta_Q \biggr)   Z_{Dj}^T Z_{Dj}\biggr)J_L + \sum \limits_{j=1}^m \lambda_j M_j, 
&\\& 
H_{cwls}(\theta_Q,\lambda) =   4\sum \limits_{j=1}^m  w_j  \biggl(\log Y_j - \log S_0 - Z_{D_j} \theta_D(L) -  \theta_Q P_j \theta_Q \biggr)  \theta_Q^T P_j^T \theta_Q P_j &\\&-2\sum \limits_{j=1}^m  w_j  \biggl(\log Y_j - \log S_0 - Z_{D_j} \theta_D(L) -  \theta_Q P_j \theta_Q \biggr)  P_j  + 2\sum\limits_{j=1}^m \lambda_j \bigl[\frac{6}{b^2}  P_j \bigr] .
\end{align*}
Then we have a sparse Hessian matrix $H_{cwls}(\Theta, \lambda) =\begin{pmatrix}
  H_{cwls}(L,\lambda) &   \\
     & H_{cwls}(\theta_Q,\lambda)  \end{pmatrix}.$

\section{Results}
The results are composed of two parts: in the first part we simulate two different datasets, conduct the estimation scheme under the proposed method and popular methods including the constrained weighted least squares (CWLS) and the traditional MLE with constraints. Finally, we reveal the performance through comparison. Part 2 is an experiment on real data from a healthy volunteer with depiction of some tensor-derived image contrasts from mean diffusivity (MD) fractional anisotropy (FA), mean kurtosis (MK), radial kurtosis $K_{\perp}$, as well as SNR ($:=S_0/\sigma$). 
\subsection{Simulation study}
The DW-MRI measurements are simulated from two models: 
1) a biexponential model of signal decay \cite{niendorf1996,maier2008}, and we can calculate the apparent diffusion and kurtosis coefficients $D_{app}$ and $K_{app}$ analytically by 
\begin{align}& \label{eq:dn}
D_{app} = f_{in}D_{in} + (1-f_{in})D_{ex}, \qquad \mbox{and} &\notag \\&
K_{app} = 3f_{in}(1-f_{in})\frac{ (D_{in} - D_{ex})^2}{D_{app}^2},
\end{align}
where $D_{in}$ and $D_{ex}$ are intracelluar and extracelluar diffusion coefficients, respectively, and $f_{in}$ is the fast diffusion relative size fraction.
2) True signal model of DKI from \eqref{eq:DKI1}, where we randomly choose certain amount of voxels from a publish data resource  {\it http://academicdepartments.musc.edu/cbi/dki/dke.html} and set the tensor parameters estimated by \cite{tabesh2011} as the ground truth, and then corrupt the simulated signals by pre-defined Rician noise. Note that we have reordered the parameters in correspondence of the design matrices $Z$ defined in Section 2.3.
SNR were chosen within the range of $[8, 40]$, and we fixed non-attenuation diffusion to be $S_0 = 1$, so that the Rician noise $\sigma = 1/\mbox{SNR}$. Then the ground truth can be analytically calculated from Eq. \eqref{eq:DKI1}.
\paragraph{\textbf{Synthetic experiment 1}}
In this experiment, we simulate two datasets. In dataset 1,
we simulated 6 voxels from six different region of interest (ROI): gray matter next to cerebration fluid (GM/CSF); gray matter next to white matter (GM/WM); thalamus (TH);  putamen and globus pallidus (PU/GP); internal capsule white matter (ICWM); frontal white matter (FWM), respectively, with the reference is to \cite{jensen2010} and also shown in Table \ref{tab:4}.
The gradient scheme contains 30 directions which were chosen to acquire the human dataset. 
Since $D_{app}$ and $K_{app}$ can be calculated analytically by Eq. \eqref{eq:dn} from \cite{jensen2010}, and according to the constraint $K_{app} \leq 3/(bD_{app})$, we calculated a maximum $b$ value $= 3532 s/mm^2$ which fits all the ROIs. In practise, however, we choose $b$ value $\leq 3000 s/mm^2$ to avoid the numerical problems which may encounter in computation. Again we took an appropriate range of $b$ values partially acquired the human dataset: 62, 249, 560, 996, 1556, 2240 $s/mm^2$. 
In dataset 2, we randomly choose 18 voxels and simulate MRI measurements from Eq. \eqref{eq:DKI1} as described above by using the same gradients and $b$ values. The aim of this experiment is to compare the performance ( e.g. the accuracy and speed ) of the WLS, CWLS, MLE methods, and in addition, we also use the CWLS method proposed by \cite{ghosh2014} using
SQP MATLAB solver, CWLS\_SQP in short, where we do not use the user-defined Hessian matrices but the default values provided by the solver in the computation. \\
\begin{minipage}{\linewidth}
\begin{center}
\captionof{table}{Ground truth (GT) of the diffusion-scalar statistics of six different ROI from dataset 1} \label{tab:1}
\begin{tabular}{|l|*{6}{c|}r}
\hline
ROI     & MD [$mm^2 /s \times 10^{-3}$]& FA & MK & $K_{\perp}$  \\
\hline
GM/CSF  &  0.9263 &0.0669 &  0.3128 & 0.4267\\
GM/WM   &  0.8595 &0.0331 &0.4748 &0.0421\\ 
TH      & 0.9371  &0.0700 &0.8204 &0.4772\\
PU/GP   & 0.7814  &0.0282&0.5449 & 0.2849\\
FWM     &  0.8351 &0.0583 & 0.6990 & 0.7052\\
ICWM    & 0.8336  &0.0193 &  0.8914 & 0.4735\\
\hline    
\end{tabular}\par
\end{center}  
\end{minipage}
\begin{minipage}{\linewidth}
\begin{center}
\captionof{table}{Variance of the diffusion-scalar statistics of dataset 1} \label{tab:2}
\begin{tabular}{|l*{6}{|c}r}
\hline
ROI   &Methods  & MD [$mm^2 /s \times 10^{-9}$]& FA [$\times 10^{-2}$] & MK & $K_{\perp}$  \\
\hline
GM/CSF  &   WLS &  0.8178 & 0.5324&0.1199& 0.4964 \\
        &   CWLS&0.4538& 0.2036& 0.7949&  0.0035\\
        &   CWLS-SQP&0.0875 & 0.7321 & 1.4916 &  0.0157\\
        &   MLE  &0.6244  & 0.1916 & 1.0574 & 0.0830\\
\hline
GM/WM  &   WLS &1.4451&1.6296&0.1937& 1.0722\\
        &   CWLS&5.0117& 0.1598&0.1598  & 0.6841\\
        &   CWLS-SQP&3.3669 & 1.6200 &0.2479 & 0.3409\\
        &   MLE  &  1.1027&  1.2616 & 0.1524 & 1.4538\\
\hline
TH  &   WLS & 0.3780&1.0613&0.0255 &0.0422\\
        &   CWLS&4.1144& 0.5919 & 0.5919 & 0.0006\\
        &   CWLS-SQP&0.9223& 0.7387 & 0.6046 & 1.3901\\
        &   MLE  &  0.0032 & 0.5170 &  0.3909 &  1.6857\\      
\hline
PU/GP &   WLS &0.0125&8.7861&0.1570& 0.2890\\
        &   CWLS&1.7518 & 7.1319&  7.1319&  0.0591\\
        &   CWLS-SQP& 0.0076 & 4.1835   & 7.6477 & 0.0012\\
        &   MLE  &  0.6757 &4.2028 & 3.8034&   0.0010\\
\hline
FWM  &   WLS & 0.0400&0.0423& 0.0075& 0.3998\\
        &   CWLS&9.5057 & 0.1406& 1.5176 & 0.2678\\
        &   CWLS-SQP &3.3476& 0.1147&2.4464 &  0.0424\\
        &   MLE  &  0.3208 & 0.0085 & 1.5472 & 0.0013\\
\hline
ICWM &   WLS &0.9640& 2.3897  & 0.6123&  0.3749 \\
        &   CWLS& 1.5569 &  2.3897&  0.0342 & 0.0099   \\
        &   CWLS-SQP& 0.0955 & 3.4572 &  0.1036 & 1.2634 \\
        &   MLE  &  0.7171 &  2.1623 & 0.0137 &  0.2415\\
\hline
\end{tabular}\par
\end{center}  
The values of the six ROIs were taken from \cite{maier2008}. 
The results are collected from dataset 1, calculating the variance between the estimates and the ground truth in Table \ref{tab:1}, and the SNR is fixed to 15. As we can see from Table \ref{tab:2}, the CWLS and the MLE methods presented in this work perform better in average than the other two, epsecially in the ROI-internal capsule white matter. The performance of WLS method is also good, especially in the ROI-frontal white matter compare with the results by the MLE, due to the low noise level of the data.
\end{minipage} 
\begin{minipage}{\linewidth}
\begin{center}
\captionof{table}{Mean square error (MSE) of the diffusion-scalar statistics from dataset 2} \label{tab:3} 
\begin{tabular}{|l|*{6}{c|}r}
\hline
Methods  & MD [$mm^2 /s \times 10^{-7}$]& FA & MK & $K_{\perp}$ &DT &KT \\
\hline
WLS  &  0.244 & 0.0134 &  0.4700 & 0.5430& 0.0564&0.7555 \\
CWLS   & 5.451 &0.0196&0.2112&0.2520&0.1071& 0.9352\\ 
CWLS-SQP & 0.392 &0.0202&0.2344 &0.9196&0.0653&0.9385\\
MLE &  0.080 &0.0125&0.0101 & 0.4533&0.0341&0.7831\\
\hline    
\end{tabular}\par
\end{center}  
\bigskip
The ground truth of MD, FA, MK, $K_{\perp}$ are on an average 1.7711 $mm^2 /s \times 10^{-3}$, 0.1334, 0.5991, 0.6160, respectively. The average computational time per voxel of CWLS, CWLS-SQP and MLE
are 4.3733, 1.4565 and 1.2908 seconds (sec.), respectively. The percentages of voxels violating constraint \#1, \#2 and \# 3 are 0, 14.81\% and 0, respectively. Again in dataset 2, we fix SNR to be 15. As can be seen the proposed MLE method performed slightly better than other methods. WLS method works also good due to low-noise level and low percentage of voxels violation
of the constraints.
\end{minipage}
\bigskip
\\
During the simulation, we fix SNR to be 15 for both datasets. From dataset 1, we can have a general understanding of information diffusion from the six different tissues by computing MD, FA, MK, and $K_{\perp}$, etc. . The results present in Table \ref{tab:1} and set as ground truth of dataset 1. Then we estimate the tensor parameters and compare the performance
by different methods from those tissues, respectively, by means of the variance and list the results in Table \ref{tab:2}. From the table, we can see that the proposed CWLS and MLE methods perform better on an average than the other two. The log-normal model works well
due to the low-noise level of this dataset. Moreover, the anisotropic level of this dataset is very weak due to the selected tissues from the reference. In dataset 2, we compute the mean square errors (MSE) of
the diffusion statistics, including the MSE of the diffusion coefficients (DT) and the diffusion kurtosis (KT) as well. We also monitor the computational time per voxel on average and the percentages of voxels violation of the constraints.
The results in Table \ref{tab:3} show us that the MLE method performs slightly better than others.

\paragraph{\textbf{Synthetic experiment 2}}
In this experiment, we generate one synthetic dataset. In dataset 3 again we use the public data resource as in the experiment 1, randomly select 180 voxels. We choose three shells with $b$ values = 500, 1000 ,1500 $s/mm^2$, and use 18 distinct gradients computed by electrostatic energy minimization algorithm which were shown to have the advantage of
maintaining the optimal coverage of the complete scan in \cite{cook2007}. The SNR is in the range [8,40] with noise ($S_0/SNR$) increasingly changing every 20 voxels to corrupt the generated signals. Again in this dataset we fix $S_0=1$.

We compare the results from WLS, CWLS with user-defined Hessian matrices (CWLS), CWLS\_SQP by using SQP MATLAB solver with default Hessian values, as well as the least squares non-linear regression method (CWLS\_LLS) by calling MATLAB function
\verb lsqnonlin  and using the initials from WLS. We also discuss the choices of good initial values in Appendix \ref{ape:3}. In order to compare the methods, we fix the estimates $S_0$ and $\sigma^2$ from WLS for all the method CWLS\_LLS, CWLS and MLE. And finally we run the who scheme of MLE method (including update $S_0$ and $\sigma^2$ ) and record the computational time.

\begin{figure}[!hbtp]\notag
\centering
\includegraphics[width=0.6\textwidth]{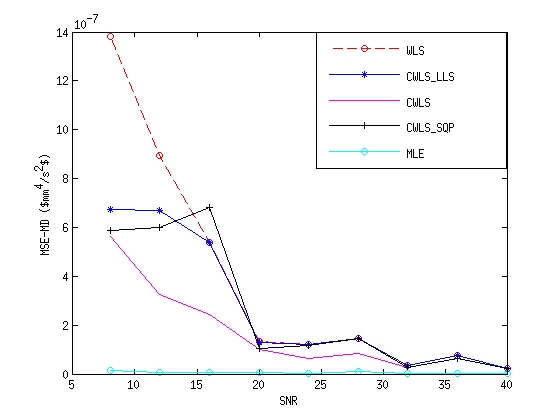}
               \centerline{a}
         \includegraphics[width=0.6\textwidth]{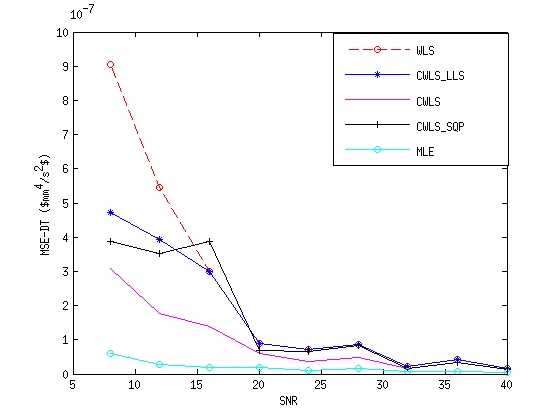}         
           \centerline{b}
\end{figure}
\newpage
\begin{figure}[!hbtp]
\centering
\includegraphics[width=0.60\textwidth]{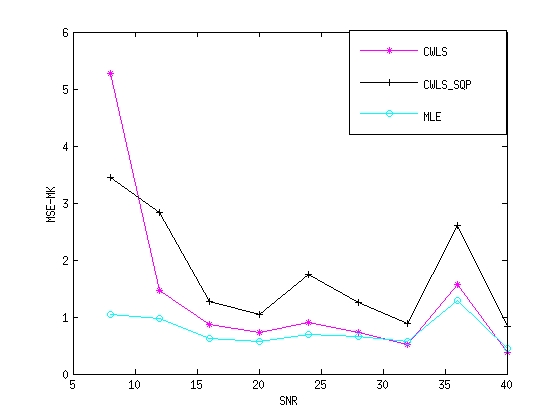}
               \centerline{c}
          \includegraphics[width=0.60\textwidth]{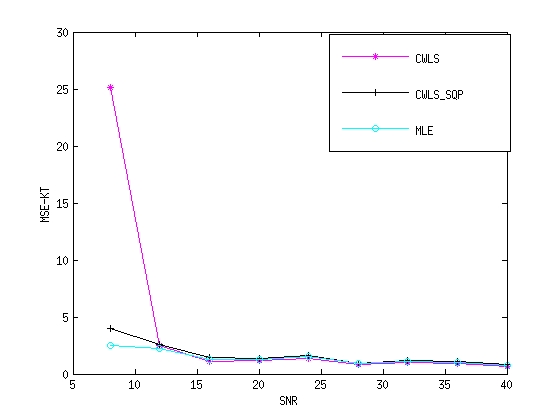}         
           \centerline{d}
    \caption{\label{fig:MSE} Mean square error (MSE) of mean diffusivity (MD, Fig. 1a), diffusion tensors (DT, Fig. 1b), mean kurtosis (MK, Fig. 1c) and kurtosis tensors (KT, Fig. 1d) from dataset 3. In Fig. 1a and 1b we compare all the methods, in Fig. 1c, d we only list the results from the CWLS, CWLS\_SQP and MLE as in the high noise range the results from the other two methods have reached out of the compared scales. }
\end{figure}
Fig. \ref{fig:MSE} shows the performance of different methods from dataset 3 by MSE of the diffusion-scalar metrics, MD in Fig.1a, DT in Fig.1b, MK in Fig. 1c and KT in Fig. 1d, respectively.
In Fig. 1a and 1b, we compare five different methods: WLS (red-break line), CWLS\_LLS (blue-star line), CWLS (cyan line), CWLS\_SQP (black-cross line) and MLE (magenta-circle line),
the color can be seen on-line. In Fig. 1c and 1d, we only compare the performance by CWLS (cyan-star line), CWLS\_SQP (black-cross line) and MLE (magenta-circle line), as in the high-noise level
the estimates by the first two methods are far away from the comparing visible region (scale). All the figures clearly indicates that our MLE method has best performance among the listed metrics. For the very high-noise data, the proposed CWLS method shows larger MSE of MK and KT than using CWLS-SQP method with out given Hessian matrices, this situation may resulted from the contribution of the Hessian matrices calculated from the log-normal signal model.
Furthermore, the percentages of voxels violating constraint \#1, \#2 and \# 3 are 7.780\% (with 14 out of 180 voxels are not satisfied positive constraint of rank-2 tensor), 55\% and 0, respectively. After comparison, we run the whole scheme of the EM-MLE method,
and monitor the running time, where the algorithm in average needs 5.1667 iterations to get convergence with 84.44\% voxels using 5 iterations.

\begin{center}
\captionof{table}{Parameters for biexponential diffusion model from normal human brains} \label{tab:4}
\begin{tabular}{l*{6}{c}r}
\hline
ROI     & $D_{in} [mm^2 /s \times 10^{-3}]$& $D_{ex} [mm^2 /s \times 10^{-3}]$ & $f_{in}$  \\
\hline
GM/CSF  & 1.479 $\pm$ 0.166 &0.466  $\pm$ 0.017 & 0.490 $\pm$ 0.012  \\
GM/WM   & 1.142 $\pm$ 0.106 & 0.338 $\pm$ 0.027 &0.622  $\pm$ 0.038\\ 
TH      & 1.320 $\pm$ 0.164 &0.271  $\pm$ 0.040 &0.617  $\pm$ 0.069  \\
PU/GP   & 1.609 $\pm$ 0.039 &0.257  $\pm$ 0.026 &0.648  $\pm$0.028 \\
FWM     & 1.155 $\pm$ 0.046 & 0.125 $\pm$ 0.026 & 0.648 $\pm$ 0.050 \\
ICWM    & 1.215 $\pm$ 0.024 & 0.183 $\pm$ 0.009 & 0.637 $\pm$ 0.020  \\
\hline
\end{tabular}\par
\end{center}  
\bigskip
The values of first six ROIs were taken from \cite{maier2008}.
\begin{center}
\captionof{table}{Optimized 18 gradient directions} \label{tab:5}
\begin{tabular}{|r|r|r|}
\hline
 0.737068&  -0.568030&  0.366160   \\
 0.795763&   0.431108&  0.425331 \\
 -0.822530&   0.367692&  0.433874 \\
 0.000650&   0.985575&  0.169239 \\
 0.228998&   0.150756&  0.961682  \\
-0.412439&  -0.753502&  0.511984 \\
-0.358616&   0.232844&  0.903979 \\
-0.891249&  -0.417614&  0.176844  \\
 0.319924&  -0.498679&  0.805586  \\
 0.309857&   0.667672&  0.676907 \\
 0.579701&  -0.807043&  -0.112374 \\
-0.209598&  -0.358489&  0.909700 \\
 0.990653&  -0.112342&  0.077367  \\
 0.153276&  -0.903274&  0.400754 \\
 0.530172&   0.845386&  0.065124  \\
-0.282930&   0.716688&  0.637423 \\
 0.720077&  -0.052737&  0.691887 \\
-0.733882&  -0.178601&  0.655377  \\
\hline
\end{tabular}\par
\end{center}  
\bigskip
This set of gradients were taken from \cite{cook2007}, point set 1, which were computed by electrostatic energy minimization algorithm and shown good performance.

\paragraph{\textbf{Summary}}
All the synthetic experiments were carried out on a 64-bit 4-core computer with 16 Gb RAM, and the CPU of each core is 3.40GHz with MATLAB.

\begin{minipage}{\linewidth}
\begin{center}
\captionof{table}{Comparison of the estimation time} \label{tal:3}
\begin{tabular}{|l|r|c|c|c|r|}
\hline
 Dataset 4  & RT, EM-MLE & Total iterations & RT, CWLS& RT, CWLS\_SQP&RT, MLE$\ast$  \\
\hline
mean   & 0.2425  & 11 &  1.4353& 0.7123 & 0.6903 \\
max   & 22.1953 &  39  &  6.7446&2.4143&3.6356 \\ 
min     &0.0205 & 2  &  0.2033 &    0.0893 &  0.0680\\
\hline
\end{tabular}\par
\end{center}  
\bigskip
The running time (RT) in average, minimum and maximum per voxel with unit second. The listed results are based on 180 voxels from six different ROIs.
The last column records the MLE method updating $\theta$ only. The MLE$\ast$ method seems to be as efficient as the CWLS\_SQP method. With the EM-MLE scheme, in some voxels we need many iterations and some others only need a few to get convergence. 
Additionally, the EM-MEL scheme may have even shorter running time on average from some small datasets than that by the MLE, because in each iteration $S_0$ and $\sigma^2$ are also updated simultaneously to obtain the optimal values, which therefore may shorten the running time when
updating $\theta$.
\end{minipage}
\bigskip

\begin{figure}[!hbtp] 
\centering
\includegraphics[width=0.7\textwidth]{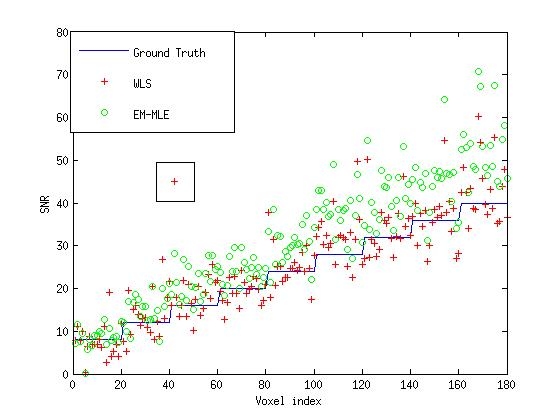}
 \caption{\label{fig:SNR} SNR of 180 voxels estimated by the WLS and the EM-MLE methods. The blue line presents the ground truth, the red cross and the green circle show the estimates by
the WLS and the EM-MLE methods, respectively.}
\end{figure}

In Fig. \ref{fig:SNR} we show the estimated SNR of 180 voxels by both the WLS and the EM-MLE methods. As can be seen that the estimates (red cross) by the WLS have large bias in the low SNR region,
and may appear some ouliers, e.g. the one marked by rectangle. In addition, they are overestimated and underestimated in the whole region of SNR. While the estimates (green circle) by 
the EN-MLE sheme performs quite well in the low SNR region. Then they fluctuate basically over the ground truth with an slight increase of deviation when the SNR is increasing. This is probably
because we set quite loose tolerances for those parameters in order to shorten the iteration of the whole running scheme, which therefore results the converged estimates have not reach the optimals.    
\subsection{Real data}
This data are part of a real experiment. It is consist of 2204 diffusion MR-images of the brain from an  healthy human volunteer,  taken from four $5mm$-thick consecutive axial slices, and measured by a Philips Achieva $3.0$ Tesla MR-scanner. The image resolution is $128\times 128$ pixels of size $1.875\times 1.875$ $mm^2$. 
 After masking  out the skull and the ventricles, 
 we remain with a region of interest (ROI) containing $18764$ voxels. In the protocol, we used all the combinations of the
$32$ gradient directions with the  $b$-values  varying in the range 0, 62, 249, 560, 996, 1556, 2240 $s/mm^2$ , with $3$ repetitions, for a  total of 7\,242\,904 data points.
\paragraph{\textbf{Results}}
In this session, we depict the results by MD Fig. 3, FA Fig. 4 as well as MK Fig. 5 from the proposed CWLS and MLE methods.The diffusion weighted MR data is in the range of (0, 581), acquired by 32 distinct gradient directions with seven different b values. 
After comparison, we can see that the image constrasts by the MLE method gain much more detailed structural information, especially in Fig.4 and Fig. 5 than those by the CWLS in the same scales.
\begin{figure}[!hbtp]\label{fig:real1}   
\centering
\includegraphics[width=0.75\textwidth]{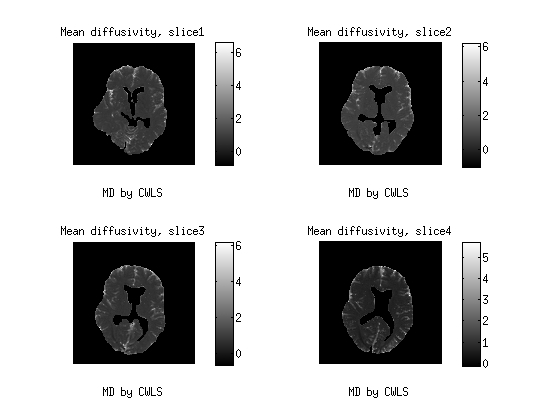}
               \centerline{a}
\end{figure} 
       \begin{figure}[!hbtp]  \notag
\centering
\includegraphics[width=0.75\textwidth]{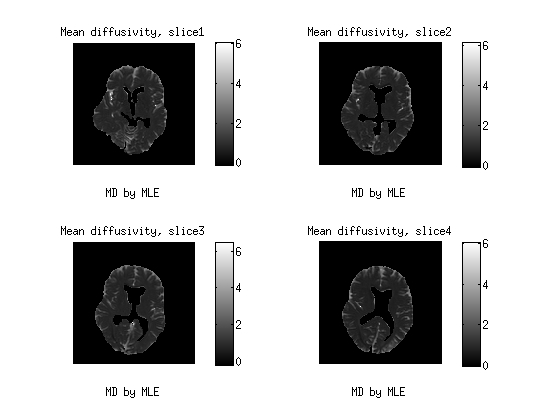}
                               \centerline{b}
\caption{\label{fig:CWLS1} 3d maps of MD by the CWLS and the MLE methods from four consecutive slices of human brain.The MD maps were scaled between (0,6) $\times 10^{-3} mm^2/s$.}
\end{figure} 
\begin{figure}[!hbtp]\label{fig:real2}    
\centering
  \includegraphics[width=0.8\textwidth]{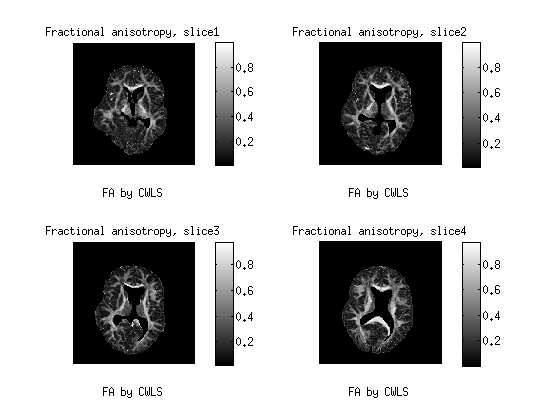}         
           \centerline{a}
      \end{figure}  
\begin{figure}[!hbtp] \notag
\centering
  \includegraphics[width=0.8\textwidth]{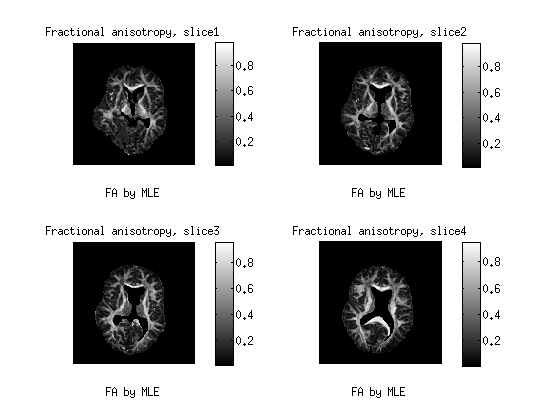}         
           \centerline{b}
 \caption{\label{fig:CWLS2} 3d maps of FA by the CWLS and the MLE methods from four consecutive slices of human brain.The FA maps is between (0,1).  }
\end{figure}  
\begin{figure}[!hbtp]\label{fig:real3}   
\centering  
\includegraphics[width=0.8\textwidth]{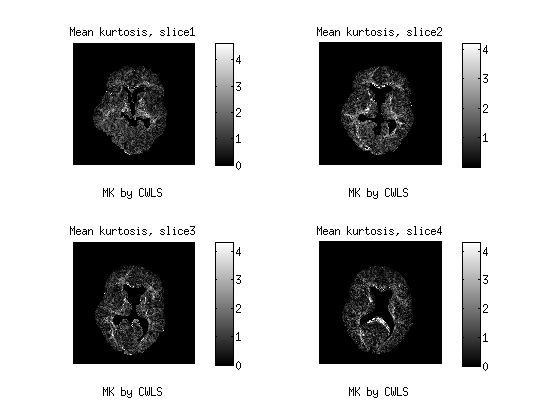}
               \centerline{a}    
       \end{figure} 
\begin{figure}[!hbtp]
\centering  
\includegraphics[width=0.8\textwidth]{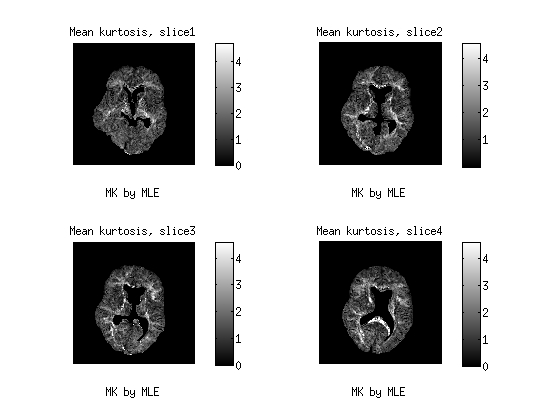}
               \centerline{b} 
\caption{\label{fig:CWLS3} 3d maps of MK by the CWLS and the MLE methods from four consecutive slices of human brain. The MK maps were scaled in the range of (0,4). }
\end{figure}    
 
\section{Discussion}
In this work, we propose an estimation scheme by the EM algorithm for MLE in contained DKI.
We use the Rician noise model of signal measurements through data augmentation to conduct the DKI estimation, which plays crucial roles at low SNRs and leads less biased estimates both in theory and what has been observed in the experiments. Using the state-of-the-art statistical methodology of data augmentation, we are able to work with a generalized linear model (GLM) of the joint likelihood derived from the Rician density. The positive constraints are imposed by Cholesky decomposition and the new parametrization of TQ for the 2nd order and kurtosis tensor, respectively. The whole scheme is not only for updating simultaneously the tensor parameters but updating the noise and the unattenuated signal. To apply this whole scheme in other simpler model such as CWLS or other DWI alternatives is straightforward. 

Using the Fisher scoring algorithm to solve optimization problem of the specific non-linear quartic regression problem from DKI, we can dramatically reduce the computational cost by deriving the gradient functions, the Hessian matrices and reducing the complexity of the Fisher information. Especially,$\theta_D$ is a function of $L$ which provide possibility to calculate the essential Fisher information for updating the parameter $L$ in the Fisher scoring method. Compare with the  
the observed information (or so-called empirical Fisher information) $\mathcal{J}(L)$, the Fisher information's algebraically simper formula, will lead substantially less computation, and it is much stable in the sense of being singular than the observed information matrix. Further details can be found in \cite{green1984}. On the other hand, the barrier method provides the possibility to impose the non-linear constraints in implementation. The two methods combined together create a constrained Fisher scoring scheme for updating the tensor parameters in DKI. Furthermore, as reported in literature that implementation of the interior point method (with the common Newton and the barrier methods) can be very difficult. In this work we prefer the use of the Fisher scoring method instead of the Newton algorithm and applying the regularization technique to smooth the Hessian matrix  by $H(\theta^{(k)}, \lambda^{(k)} + \|\S(\theta^{(k)})\|)$ before calculating the Fisher information. As a consequence, the results show us that  our constrained Fisher scoring scheme works very efficiently.

\section{Acknowledgement}
The author would like to thank emeritus professor Antti Penttinen from University of Jyv\"askyl\"a carefully read the manuscript and made insightful comments, in which this work can be well represented. Moreover, acknowledge Dario Gasbarra for useful discussion and Dr. Juha Raivola for the real data contribution.
This work was funded by Doctoral Program in Computing and Mathematical Sciences (COMAS) and Department of Mathematics and Statistics, University of Jyv\"askyl\"a.

\appendix
\renewcommand{\thesubsection}{\Alph{subsection}}
\numberwithin{equation}{subsection}
\renewcommand{\theequation}{\Alph{subsection}.\arabic{equation}}
   
\section*{Appendix}

\section{Fisher scoring method for $L$}\label{ape:1}
Let's ${\bf \zeta}_j^{(k)} =   \exp(  Z_{Dj} \theta_D ^{(k)}), {\bf \psi}_j^{(k)} = \exp \biggl(     (\theta_Q^{(k)})^T P_j   \theta_Q^{(k)} \biggr )$ and ${\bf\tau}_j^{(k)} =  Y_j \bigl\langle \cos( \varphi_j)\bigr \rangle^{(k)} $.
\\
The score is of $\theta_D$ is the first derivative of Eq. \eqref{eq:objective} w.r.t. $\theta_D$ given by
\begin{align}
\bigtriangledown q(\theta_D) =   (\sigma^{-2})^{(k)} \sum \limits_{j=1}^m \biggl\{ (S_0^{(k)})^2    ({\bf \zeta}_j^{(k)})^2 ({\bf\psi}_j^{(k)})^2  Z_D^T  - S_0^{(k)} {\bf \tau}_j^{(k)} {\bf \zeta}_j^{(k)} {\bf \psi}_j^{(k)}Z_D^T \biggr\},
\end{align}
and the Hessian matrix is
\begin{align}&\label{eq:Hessian_theta}
\bigtriangledown^2 q(\theta_D) =    (\sigma^{-2})^{(k)}\sum \limits_{j=1}^m \biggl\{ 2 (S_0^{(k)})^2  ({\bf \zeta}_j^{(k)})^2 ({\bf\psi}_j^{(k)})^2  Z_D^TZ_D - S_0^{(k)} {\bf \tau}_j^{(k)} {\bf \zeta}_j^{(k)} {\bf \psi}_j^{(k)} Z_D^TZ_D \biggr\},
\end{align}
and observed information $\mathcal{J}(\theta_D)= -\bigtriangledown^2 q(\theta_D)$ is defined as the minus Hessian.
\\
The score of $L$ expresses
\begin{align}
\bigtriangledown q(L) =     (\sigma^{-2})^{(k)} \sum \limits_{j=1}^m \biggl\{ (S_0^{(k)})^2    ({\bf \zeta}_j^{(k)})^2 ({\bf\psi}_j^{(k)})^2  Z_D^T J_L - S_0^{(k)} {\bf \tau}_j^{(k)} {\bf \zeta}_j^{(k)} {\bf \psi}_j^{(k)} Z_D^T J_L\biggr\},
\end{align}
and the corresponding Hessian matrix is
\begin{align}&\label{eq:Hessian_L}
\bigtriangledown^2 q(L) =  J_L^T (\bigtriangledown^2 q(\theta)) J_L +  \bigtriangledown q(\theta_D) \frac{\partial^2  \theta_D(L)}{\partial L_{k} \partial L_{h} } &\\&
=    (\sigma^{-2})^{(k)} \sum \limits_{j=1}^m \biggl\{ J_L^T\biggl( 2(S_0^{(k)})^2  ({\bf \zeta}_j^{(k)})^2 ({\bf\psi}_j^{(k)})^2 Z_{Dj}^TZ_{Dj} \notag -  S_0^{(k)}{\bf \tau}_j^{(k)}  {\bf \zeta}_j^{(k)} {\bf \psi}_j^{(k)}  Z_{Dj}^TZ_{Dj}\biggr) J_L\biggr\} 
 &\\&-(\sigma^{-2})^{(k)} \sum_{j=1}^m  \biggl\{ \biggl( (S_0^{(k)})^2  ({\bf \zeta}_j^{(k)})^2 ({\bf\psi}_j^{(k)})^2 
 - S_0^{(k)} {\bf \tau}_j^{(k)} {\bf \zeta}_j^{(k)} {\bf \psi}_j^{(k)}  \biggr)M_j \biggr\}.  
\end{align}
where
\begin{align*}&  
M_j = Z_{Dj} \frac{\partial^2 \theta_{D}(L)}{\partial L_{k} \partial L_{h} } =  \begin{pmatrix}
  2 Z_{1j} &      &  &  Z_{4j} &  Z_{5j} &   \\
       & 2 Z_{2j} &  &   &  & Z_{6j} \\
       &      &2 Z_{3j} & & &   \\
       &Z_{4j} &   &2 Z_{2j} &&Z_{6j}\\
     Z_{5j}   & Z_{6j}&   &  &2 Z_{3j}& \\
      &  &   & Z_{6j}& & 2 Z_{3j}
\end{pmatrix}.
\end{align*} 

The Fisher information is given by
\begin{align*}& 
\bigl\langle  \mathcal{J}(L)^{(k)} \bigr\rangle := \E \bigl[ -\bigtriangledown^2 \log \pi(y; \theta_D(L)) \bigr]=\notag &\\&
-(\sigma^{-2})^{(k)}\sum\limits_{j=1}^m \biggl\{ J_L^T\biggl( 2 (S_0^{(k)})^2    ({\bf \zeta}_j^{(k)})^2 ({\bf\psi}_j^{(k)})^2  Z_{Dj}^TZ_{Dj} -  S_0^{(k)}   {\bf \tau}_j^{(k)} {\bf \zeta}_j^{(k)} {\bf \psi}_j^{(k)}   Z_{Dj}^TZ_{Dj}\biggr) J_L\biggr\},&\\&
\mbox{with the expectation at $\tilde \theta_D$, the current value of $\theta_D$},&\\&
\E\bigl[\bigtriangledown q(\theta_D) \bigr] =0 \qquad \mbox{and} &\\&
\E\bigl[\bigtriangledown^2 q(\theta_D)\bigr]=  (\sigma^{-2})^{(k)} \sum \limits_{j=1}^m \biggl\{ (S_0^{(k)})^2    ({\bf \zeta}_j^{(k)})^2 ({\bf\psi}_j^{(k)})^2  Z_D^T  - S_0^{(k)} {\bf \tau}_j^{(k)} {\bf \zeta}_j^{(k)} {\bf \psi}_j^{(k)}Z_D^T \biggr\}.
\end{align*} 
Note that $\theta_D$ is a function of $L$ which provide possibility to calculate the essential Fisher information equalling to the expectation value of (or minus) Hessian matrix for updating $L$ in the Fisher scoring method. Compared with the  
the observed information (or so-called empirical Fisher information) $\mathcal{J}(L)$, the Fisher information's algebraically simper formula, will lead substantially less computation, and it is much more stable in the sense of being singular than the observed information matrix. Details can be found in \cite{green1984}.

 \section{Constrained Fisher scoring method}\label{ape:2} 
Using the barrier method we form two Lagrangian of problems presented in Eq. \eqref{eq:sysD} and Eq. \ref{sys:objectiveQ}. Firstly, we need compute the score $\S(\cdot)$ and set its components to be zero to find the necessary conditions of the optimal:
\begin{align*}&
\S_{\theta} =    \bigtriangledown f(\theta) + A(\theta)^{\top}  \lambda  = 0, &\\&
\S_{\nu} = <\lambda, \nu> = \mu, &\\&
\S_{\lambda}= g(\theta) + \nu = 0,
\end{align*}
where $<\cdot>$ is an operator of inner product, and $A(\theta) := \bigtriangledown g(\theta)$ with dimension $ d\times m$. In particular, we see why the barrier function is used in the logarithmic form.
Then we need compute the Hessian matrix 
$$H(\theta, \nu, \lambda) =  \begin{pmatrix}
  H(\theta, \lambda) &  0    &  A(\theta)^{\top}  \\
     0  & diag(\lambda) & diag(\nu)  \\
        
     A(\theta) & \I_{m\times 1}  & 0
\end{pmatrix},$$
where $diag(\cdot)$ is diagonalizing operator to construct the vector to be $m\times m$ matrix.
Applying the Fisher scoring method, we update 
\begin{align*}&\label{eq:fisher-score}
\theta^{(k+1)} \leftarrow   \theta^{(k)} +\alpha \biggl(  \mathcal{I}_\theta  \biggr)^{-1} \S_{\theta}  , &\\&
 \lambda^{(k+1)} \leftarrow   \lambda^{(k)} +\beta \biggl(  \mathcal{I}_\lambda  \biggr)^{-1} \S_{\lambda}  , \; &\\&
\mbox{with} \qquad \mathcal{I}_\theta = H(\theta, \lambda), \qquad  \mbox{or some regularized form, e.g. mentioned in this work.} &\\&
\S_\theta =   A(\theta)^{\top}  (\lambda^{(k+1)} - \lambda^{(k)} ) - \bigtriangledown f(\theta) +  A(\theta)^{\top}   \lambda &\\&
\mbox{and} \qquad
\mathcal{I}_\lambda = diag(\nu)/ diag(\lambda), \qquad
\S_\lambda =  A(\theta) (\theta^{(k+1)} - \theta^{(k)} ) + g(\theta) + \mu / \lambda ,
\end{align*} 
where $\alpha$ is a positive  primal parameter, and 
$\beta$ is a positive dual step parameter. To improve the convergence of the algorithm, the step parameters can be iteratively reduced by monitoring the logarithmic likelihood \cite{jorgensen}. 
This is the so-called the Levengerg-Marquart algorithm. Beside the barrier parameter $\mu$ should be decrease as well during the iteration. All the above can be achieved by calling MATLAB optimization toolbox, function \verb fmincon  with the interior point method (IP). However in practice, for IP method can be very difficult to implement, if the selection of regularization technique, the step parameters, and the barrier parameter are not mutually consistent. In the sense, this algorithm requires skilful designs from users, including
calculation of Hessian matrices, choices of regularization, choices of stopping criteria of step parameters regarding to a specific problem in order to make the algorithm works efficiently.

\section{Choices of good initial values}\label{ape:3}
In this section, we discuss a possible solution to obtain {\it good} initial values of the tensor parameters fulfilling the positive constraints for saving the computational cost.

Firstly, we can use the DTI approach to estimate the 2nd-order diffusion tensor, and then apply Cholesky decomposition to get the initials of $L$. When encountering non-positive definite diffusion tensor matrices ($D, ~3\times 3$), we can set the corresponding non-positive eigenvalues to be negligible and positive. In such a way, we gain positive definite $D$ and preserve the directions of positive curvature in the original tensor matrices.

In order to get good initial values for $Q$, we can call the Kurtosis model presented in Eq. \eqref{eq:DKI1} and calculate the kurtosis tensor $\theta_W$, and then construct the $6\times 6$ Gram matrix ($G$) by the fifteen distinct elements in a 4th-order tensor matrix, denoted by $W$ presented as a matrix from the tensor parameter $\theta_W$. Since $G$ are symmetric, we can define $W \rightarrow G = \begin{pmatrix}
  M &   N \\
   N    & S       
\end{pmatrix}$
with
$N = \begin{pmatrix}
\frac 1 2 W_{1112}  &\frac 1 2  W_{1113}&  d \\
\frac 1 2 W_{1222}  & e&   \frac 1 2 W_{2223}\\
 f  & \frac 1 2  W_{1333}& \frac 1 2 W_{2333} \\
\end{pmatrix},$ 
$Q$ can possibly be obtained by solving the system of equations
\[
T A = A T = N, \] 
where we apply the QR decomposition w.r.t matrix $N$ to reformulate $Q$ as
$Q = \begin{pmatrix}
 T \\
 A       
\end{pmatrix},$ with two $3\times 3$ matrices, in particular, $T$ are lower triangular matrices, 
and some choices of parameters $d,~ e,~f$ in the Gram matrix can be found in \cite{basser2003} in order to make the rank of $G$ equal to 3. Note that in such reformation, each $Q$ contains the same number of distinct entries as $W$, i.e. fifteen instead of eighteen. The detailed interpretation can be found in \cite{barmpoutis2007}. The above scheme can be conducted by the least squares(LS) and the weighted least squares WLS) methods without constraints, simultaneously, we get $S_0$ and the noise parameter $\sigma^2$ at each voxel.
\end{document}